\documentstyle[12pt]{article}

\font\titlefont=cmbx10 scaled \magstep2
\begin{document}

\begin{flushright}
\vspace*{-2cm}
hep-th/9408172  \\ TUTP-94-14  \\ August 1994
\vspace*{1cm}
\end{flushright}

\begin{center}
{\titlefont DECOHERENCE AND VACUUM FLUCTUATIONS }
\vskip .4in
L.H. Ford \\
\vskip .1in
Institute of Cosmology\\
Department of Physics and Astronomy\\
Tufts University\\
Medford, Massachusetts 02155\\
\end{center}

\vskip .4in

\centerline{\bf Abstract}
\vskip 0.2in
\baselineskip=12pt

The interference pattern of coherent
electrons is effected by coupling to the quantized electromagnetic field.
The amplitudes of the interference maxima are changed by a factor which
depends upon a double line integral of the photon two-point function
around the closed path of the electrons.
The interference pattern is sensitive to shifts in the vacuum
fluctuations in regions from which the electrons are excluded. Thus this
effect combines aspects of both the Casimir and the Aharonov-Bohm effects.
The coupling to the quantized electromagnetic field tends to decrease
the amplitude
of the interference oscillations, and hence is a form of decoherence. The
contributions due to photon emission and to vacuum fluctuations may be
separately identified. It is to be expected that photon emission leads to
decoherence, as it can reveal which path an electron takes. It is less
obvious that vacuum fluctuations also can cause decoherence. What is directly
observable is a shift in the fluctuations due, for example, to the presence
of a conducting
plate. In the case of electrons moving parallel to conducting
boundaries, the dominant decohering influence is that of the vacuum
fluctuations. The shift in the interference amplitudes can be of the order
of a few percent, so experimental verification of this effect may be possible.
The possibility of using this effect to probe the interior of matter, e.g.,
to determine the electrical conductivity of a rod by means of electrons
encircling it is discussed. The effect of squeezed states of the photon field
are considered, and it is shown that such states may either enhance or suppress
the decohering effects of the vacuum fluctuations.
\vskip .7in
\centerline{ \hrulefill}
\vskip .1in
To be published in the Proceedings of the Conference on Fundamental Problems
in Quantum Theory, University of Maryland, Baltimore County, June 18-22, 1994.
\newpage

\centerline{\bf Vacuum Fluctuations and Photon Emission}
\vskip .1in
\baselineskip=14pt

       It is well known that the interaction of a quantum system with its
environment can destroy quantum coherence. In this paper, a particular
example of this phenomenon will be discussed: the coupling of coherent
electrons to the quantized radiation field. This coupling gives rise both
to the possibility of photon emission and of interaction of the electrons
with the electromagnetic vacuum fluctuations. Consider an electron
interference experiment in which coherent
electrons may travel from $x_i$ to $x_f$ along either of two classical
paths, $C_1$ or $C_2$. First let us recall the analysis of this experiment
when the effects of the electromagnetic field are ignored. Let $\psi_1$
and $\psi_2$ be the amplitudes for an electron to travel along $C_1$ and $C_2$,
respectively. Then the superposed amplitude is $\psi = \psi_1 + \psi_2$,
and the number density of electrons detected at $x_f$ is
\begin{equation}
n_0 (x_f) = |\psi|^2 = |\psi_1|^2 +|\psi_2|^2 +2 Re (\psi_1 {\psi_2}^*)\,,
                                          \label{eq:numdens0}
\end{equation}
the last term being responsible for the interference pattern. Note that
$C_1$ and $C_2$ are {\it spacetime} paths, as the events of emission
and detection of the electrons occur at different times as well as different
points in space.

      We now wish to couple the electrons to the quantized electromagnetic
field and examine the effect upon the interference pattern. This problem has
been analyzed in detail in Ref. \cite{F93}. Here we will quote and discuss
some of the main results. When both photon emission and vacuum fluctuation
effects are present, the number density of electrons detected at $x_f$
becomes
\begin{equation}
n(x_f) = |\psi_1|^2 +|\psi_2|^2 +
                2e^W Re (e^{i\phi}\psi_1 {\psi_2}^*)\, . \label{eq:numdens1}
\end{equation}
Here $\phi$ is a phase shift introduced by the interaction, and $W$ is a
function which describes the change in the amplitude of the interference
oscillations (the contrast). The phase shift $\phi$ includes the
Aharonov-Bohm shift due to any classical electromagnetic fields generated
by the electrons. However, in this paper we will be primarily concerned
with the amplitude of the interference oscillations, as this quantity carries
the information about the vacuum fluctuations.
The explicit form for $W$ is
\begin{equation}
W = -2\pi\alpha \oint_{C}d{x_\mu}\oint_{C}d{{x'}_\nu} D^{\mu\nu}(x,x'),
      \label{eq:defw}
\end{equation}
where $\alpha$ is the fine structure constant, and $C = C_1 - C_2$ is the
closed spacetime path obtained by traversing $C_1$
in the forward direction and $C_2$ in the backward direction. Note that
Eq. (~\ref{eq:defw}) is independent of the direction in which the integrals
around $C$ are taken.
The photon Hadamard (anticommutator) function, $D^{\mu\nu}(x,x')$, is defined
by
\begin{equation}
D^{\mu\nu}(x,x') =  {1 \over 2} \langle 0| \{ A^\mu (x), A^\nu (x') \}
                     |0 \rangle \,. \label{eq:twopoint}
\end{equation}
   By means of the four dimensional Stokes theorem, we may write
\begin{equation}
W = -2\pi\alpha \int d{a_{\mu\nu}}\int d{{a'}_{\rho\sigma}}
      D^{\mu\nu;\rho\sigma}(x,x'),  \label{eq:surfaceint}
\end{equation}
where $d{a_{\mu\nu}}$ is the area element of the timelike two-surface enclosed
by $C$, and
\begin{equation}
D^{\mu\nu;\rho\sigma}(x,x') = {1 \over 2} \langle 0| \{ F^{\mu\nu} (x),
                     F^{\rho\sigma} (x') \} |0 \rangle
\end{equation}
is the Hadamard function for the field strengths.
Equation (~\ref{eq:surfaceint}) has the remarkable interpretation that the
electrons are sensitive to vacuum fluctuations in regions from which they are
excluded. This is analogous to the situation in the Aharonov-Bohm
effect\cite{AB},
where the phase shift can depend upon classical electromagnetic fields in
regions which the electrons cannot penetrate. The spacetime geometry of
the paths $C_1$ and $C_2$ encircling a region is illustrated in Fig. 1.

      In Eq. (~\ref{eq:numdens1}), it was assumed that both photon emission
and vacuum fluctuation effects are present at the same time. This is the
usual case in an interference experiment in which no attempt is made to
detect the emitted photons. However, in principle, it is possible to
distinguish
the two by means of a {\it veto} experiment. Suppose that we arrange for the
flux of electrons to be sufficiently low that only one electron is in the
apparatus at any one time, and that any photons emitted be detected. Whenever
a photon is in fact detected, the electron counters are switched off for a
sufficient time to insure that the associated electron is not counted. In this
way, we guarantee that the interference pattern is comprised only of those
electrons which have not emitted photons. In this case, the relevant contrast
factor is no longer $e^W$, but rather $e^{W_F}$, where
\begin{equation}
W_F = -2\pi\alpha \biggl(\int_{C_1}d{x_\mu}\int_{C_1}d{{x'}_\nu} +
\int_{C_2}d{x_\mu}\int_{C_2}d{{x'}_\nu}\biggl) D^{\mu\nu}(x,x')\,.
      \label{eq:defwf}
\end{equation}
This function describes the effect of the vacuum fluctuations. Similarly,
the effects of photon emission are described by the function
\begin{equation}
W_\gamma = W -W_F = 2\pi\alpha \biggl(\int_{C_1}d{x_\mu}\int_{C_2}d{{x'}_\nu} +
\int_{C_2}d{x_\mu}\int_{C_1}d{{x'}_\nu}\biggl) D^{\mu\nu}(x,x')\,.
      \label{eq:defwgamma}
\end{equation}
Note that the vacuum fluctuation effect involves a double line integral over
each path separately, whereas the photon emission contribution is a cross term
involving a line integral over each path.
\vskip .3in

\centerline{\bf Effect of a Conducting Plate}
\vskip .1in

     The line integrals in Eqs. (~\ref{eq:defw}) and (~\ref{eq:defwf}) are
divergent due to the short distance singularity of $D^{\mu\nu}(x,x')$
when $x \rightarrow x'$. This is one of the familiar ultraviolet divergences
in quantum field theory. A simple way to avoid dealing with this divergence
is to consider only changes in $W$ or $W_F$ due to an external influence,
such as the presence of a conducting boundary. In this case, we replace
$D^{\mu\nu}(x,x')$ by $D_R^{\mu\nu}(x,x')$, the renormalized Hadamard function
obtained by subtracting the empty space function. Let $W_R$ be the function
obtained by making this replacement in Eq. (~\ref{eq:defw}). As compared to
the empty space case, the amplitude of the interference oscillations is
multiplied by the factor $e^{W_R}$.
 The interference pattern
can now become sensitive to shifts in the vacuum fluctuations, including
shifts occurring in excluded regions. In this case, one has an effect which
combines aspects of both the Aharonov-Bohm and the Casimir
effects\cite{Casimir}. The simplest geometry in which the effects of vacuum
fluctuations may be investigated is where one of the electron paths skims
above a perfectly conducting plate. Suppose that path $C_1$ travels for
a distance $L$ at a height $z$ above the conducting plate, and later recombines
with path $C_2$, which travels far away from any conductors. Let $v$ be the
speed of the electrons. In the limit that $Lc/v \gg z$, i.e. the electron's
flight time over the plate is long compared to the light travel time to
the plate, we have
\begin{equation}
W_R \approx - {\alpha \over {\pi}}
   \Bigl[ 1 + log \Bigl({Lc \over {2v z}} \Bigr)\Bigr]\,.  \label{eq:WR1}
 \end{equation}
Note that this is the small $z$ approximation to a function which vanishes
in the limit that $z \rightarrow \infty$, as required by the fact that
$W_R=0$ in empty space. We first observe that
\begin{equation}
W_R < 0 \,.
\end{equation}
This means that the shift in the vacuum fluctuations due to the plate causes
a decrease in the contrast, and hence there has been a loss of quantum
coherence.

      That photon emission can cause decoherence is no surprize. The detection
of sufficiently short wavelength photons could reveal which path the electron
has taken. It is perhaps less obvious that vacuum fluctuations are also
capable of causing decoherence. One might intuitively think of the electrons
as being subjected to random force fluctuations which eventually lead to
decoherence. Although in the above example $W_R < 0$, there is no reason in
principle why one could not have $W_R > 0$, in which case the presence of
the conducting boundary would suppress the decohering effects of the vacuum
fluctuations.
\vskip .3in

\centerline{\bf Squeezed States of the Radiation Field}
\vskip .1in

     Such a situation may be displayed explicitly when one replaces the
boundary by photons in a squeezed vacuum state. Let the photon field be
in squeezed vacuum state for a single mode, which can be defined
as\cite{Caves81}
\begin{equation}
|\zeta\rangle=S(\zeta)\,|0\rangle = \exp[{1\over 2}\zeta^\ast a^2
 -{1\over 2}\zeta ({a^\dagger})^2]\,|0\rangle \,,
\end{equation}
where $\zeta = re^{i\delta}$ is an arbitrary complex number.
It may be shown that, in this state,
\begin{equation}
\langle a^\dagger a \rangle = \sinh^2 r \, ,
\end{equation}
and
\begin{equation}
\langle a^2 \rangle = -e^{i\delta} \sinh r \cosh r \, .
\end{equation}
We wish to calculate the shift in $W$ from the actual vacuum state. This
shift, $W_R$, is given by Eqs. (~\ref{eq:defw}) and (~\ref{eq:twopoint})
with the expectation value taken in the state $|\zeta\rangle$. The photon
operator product is now understood to be
normal ordered with respect to the vacuum state. The result is
\begin{equation}
W_R = -4\pi \alpha |\eta|^2 \sinh r \,
                         [\sinh r +\sin(2\theta+\delta)\cosh r]\,,
\end{equation}
where
\begin{equation}
\eta = |\eta|e^{i\theta} = \oint_{C}d{x_\mu} f^{\mu}(x) \,,
\end{equation}
and $f^{\mu}(x)$ is the mode function of the excited mode.
The key point is that one can arrange for $W_R$ to have either sign by
an appropriate choice of the state parameter $\delta$. Thus the effect of
the squeezed state can be either to enhance or suppress decoherence.
In the above example, both vacuum fluctuation and photon emission contributions
were included. However, one can treat them separately and show that the
shifts in each of $W_V$ and $W_\gamma$ can have either sign. That squeezed
states can suppress the effects of quantum fluctuations below the vacuum
level is well known. A squeezed vacuum state necessarily has the expectation
value of the energy density negative in certain regions as a result of this
suppression. A phenomenon somewhat analogous to the present situation
arises when one considers the coupling of a spin system in a classical
magnetic field to the electromagnetic vacuum fluctuations. These fluctuations
tend to cause a depolarization of the system. Photons in a squeezed state
can temporarily reduce this effect, and cause the mean magnetic moment of
the system to {\it increase} relative to the vacuum value\cite{FGO}.
\vskip .3in

\centerline{\bf Probing the Interior of Matter}
\vskip .1in

     Let us now return to the issue of nonlocality, the ability of electrons
to serve as remote probes. A simple illustration of this would arise in the
following experiment: send electrons around either side of a cylinder of
radius $R$ filled with a material of finite electrical conductivity, as
illustrated in Fig. 2. Also
arrange that the outer wall of the cylinder is made of a material of very
high conductivity. This wall both excludes the electrons and insures that
the renormalized field strength Hadamard
function, $D^{\mu\nu;\rho\sigma}_R(x,x')$, outside the cylinder is independent
of the material on the interior. However, the change in the interference
pattern contrast due to the presence of the cylinder also depends upon
$D^{\mu\nu;\rho\sigma}_R(x,x')$ inside the cylinder, and hence upon the
conductivity of the material on the interior.
An explicit calculation of $D^{\mu\nu;\rho\sigma}_R(x,x')$ in the interior
requires a knowledge of the dielectric function $\epsilon(\omega)$
of the metal in question. However, we may make an order of magnitude
estimate without this detailed information.
    We may adapt Eq. (~\ref{eq:WR1}) to obtain an estimate of $W_R$ for the
present situation. Let $\lambda_P$ be the wavelength associated with
the plasma frequency in the metal in question. It is essentially the
cutoff wavelength, in that only modes whose wavelengths are of this
order or longer are effected by the interior material. It plays a role
analogous to $z$ in Eq. (~\ref{eq:WR1}). Our
estimate will be based upon the assumption that a result of the form
of Eq. (\ref{eq:WR1}) also holds in the present situation. If we set
$z \approx \lambda_P$ and $L \approx R$, the radius of the rod,
our estimate for $W_R$ may be written as
\begin{equation}
W_R \approx - {10^{-3}} ln \Bigl({{cR} \over {v \lambda_P }} \Bigr),
         \qquad R \gg \lambda_P, \qquad v \ll c.
                                               \label{eq:WR2}
 \end{equation}
The result is only weakly (logarithmically) dependent upon the cutoff,
$\lambda_P$. For example, if $R= 1cm$, $\lambda_P = 810 \AA$ (the
approximate plasma wavelength of aluminum), and $v = 0.1c$ (corresponding
to $2.5 \,keV$ electrons), we find that $W_R \approx - 10^{-2}$.
Because the amplitude of the interference oscillations is proportional
to $e^{W_R} \approx 1 + W_R$, we see that the effect of the presence
of the rod is to decrease this amplitude by about $1\%$ in this example.
If one were to use a
material with a different plasma wavelength, $\lambda'_P$, the amplitude
shift will now be $W'_R = W_R + \Delta W_R$, where
\begin{equation}
\Delta W_R \approx  {10^{-3}} ln(\lambda'_P/\lambda_P).
\end{equation}
This change will typically be of the order of a few times $10^{-4}$ of
the interference oscillation amplitude. Thus measuring
the electron interference pattern to this accuracy would enable one to
ascertain whether the material in the rod is, for example, aluminum or
magnesium ($\lambda'_P = 1170 \AA$).
The effect which we have estimated is primarily due to the effects
of the vacuum fluctuations, rather than photon emission, as the magnitudes
of the integrals in Eq. (~\ref{eq:defwgamma}) are determined by the
separation of the paths, which is at least as large as $R$.
\vskip .3in

\centerline{\bf Experimental Prospects}
\vskip .1in

      Let us conclude with some remarks on the possibility of experimentally
verifying the effects discussed in this paper. The effect discussed in
the previous
paragraph seems to be large enough to be possibly measureable. However,
in this case,
the theoretical prediction, Eq. (~\ref{eq:WR2}), is only an order of
magnitude estimate. For the case of electrons moving parallel to a
conducting plate, the theory is more clear cut. In principle, $W_R$ as
given by Eq. (~\ref{eq:WR1}) could
be made arbitrarily large if one could make $L$ large and $z$ and $v$
small. In an actual experiment, the image charge force experienced by
an electron will limit the possible range of parameters. In particular,
the image effect is negligible so long as
\begin{equation}
 L \ll (6 cm)\biggl(\frac{v}{c}\biggr)
                       {\biggl({{z} \over {1\mu}}\biggr)}^{3\over 2}\,.
                                                 \label{eq:image}
\end{equation}
It is not difficult to satisfy Eq. (~\ref{eq:image}) with choices of
parameters that yield a contrast shift, $|W_R|$, of the order of $1\%$.
For example, if we again let $v = 0.1c$ and take $L= 1\,m$ and $z=1\,mm$,
then Eq. (~\ref{eq:image}) is satisfied, and we have $|W_R|\approx 2.2\%$.
Thus detection of the vacuum decoherence effect seems to be a possibility.

\vskip .4cm
{\bf Acknowledgements:}  This work was
supported in part by National Science Foundation Grant No. PHY-9208805.

\section*{Figure Captions}
\begin{itemize}

\item{[1]} Electrons traverse spacetime paths $C_1$ and  $C_2$ to reach
point $({\bf x},t)$. The interference pattern depends upon vacuum
fluctuations in the crosshatched region in
the interior of these paths, which is the domain of integration in
Eq. ~\ref{eq:surfaceint}. The cylinder is the world history
of an object contained  within the electron paths.
The spatial projections of the spacetime paths $C_1$ and  $C_2$ onto
the $xy$ plane are $C'_1$ and  $C'_2$, respectively. The shaded region
in this plane is the cross section in the $xy$ plane of the object around
which the electrons travel.

\item{[2]} A ``coaxial cable'' consisting of an outer cylinder which
prevents electrons from penetrating to the interior, and an inner rod,
with radius $R$, of material of finite conductivity.

\end{itemize}

\end{document}